\documentclass[twocolumn]{aastex63}

\received{January 8, 2021}

\submitjournal{AJ}

\shorttitle{The Stars Kepler Missed}
\shortauthors{Wolniewicz et al.}

\begin{document}

\title{The Stars Kepler Missed: Investigating the Kepler Target Selection Function Using Gaia DR2}

\author[0000-0002-2087-1634]{Linnea M. Wolniewicz}
\affiliation{Department of Astrophysical and Planetary Sciences, University of Colorado, 2000 Colorado Ave, Boulder, CO 80305, USA}

\author[0000-0002-2580-3614]{Travis A. Berger}
\affiliation{Institute for Astronomy, University of Hawai`i, 2680 Woodlawn Drive, Honolulu, HI 96822, USA}

\author[0000-0001-8832-4488]{Daniel Huber}
\affiliation{Institute for Astronomy, University of Hawai`i, 2680 Woodlawn Drive, Honolulu, HI 96822, USA}

\begin{abstract}

The Kepler Mission revolutionized exoplanet science and stellar astrophysics by obtaining highly precise photometry of over 200,000 stars over 4 years. A critical piece of information to exploit Kepler data is its selection function, since all targets had to be selected from a sample of half a million stars on the Kepler CCDs using limited information. Here we use Gaia DR2 to reconstruct the Kepler selection function and explore possible biases with respect to evolutionary state, stellar multiplicity, and kinematics. We find that the Kepler target selection is nearly complete for stars brighter than $Kp < 14$ mag and was effective at selecting main-sequence stars, with the fraction of observed stars decreasing from 95\% to 60\% between $14 < Kp < 16$ mag. We find that the observed fraction for subgiant stars is only 10\% lower, confirming that a significant number of subgiants selected for observation were believed to be main-sequence stars. Conversely we find a strong selection bias against low-luminosity red giant stars ($R \approx 3-5 R_\odot$, $T_{eff} \approx  5500$K), dropping from 90\% at $Kp = 14$ mag to below 30\% at $Kp = 16$ mag, confirming that the target selection was efficient at distinguishing dwarfs from giants. We compare the Gaia Re-normalized Unit Weight Error (RUWE) values of the observed and non-observed main sequence stars and find a difference in elevated ($>$ 1.2) RUWE values at $\sim\,5\,\sigma$ significance, suggesting that the Kepler target selection shows some bias against either close or wide binaries. We furthermore use the Gaia proper motions to show that the Kepler selection function was unbiased with respect to kinematics. 

\end{abstract}

\section{Introduction}
\label{s.Introduction}

The Kepler mission \citep{Borucki2010}, officially retired in 2018, has left behind a legacy dataset for stellar astrophysics and exoplanet science. One of the biggest breakthroughs enabled by Kepler was our understanding of exoplanet occurrence rates as a function of planet size, orbital period and stellar type. For example, many planets observed around Kepler host stars have been found to have sizes between Earth and Neptune \citep{Howard2012}, a population that is absent in our own solar system. \citet{Dressing2013} found that for the M dwarf stars in the Kepler sample, the Earth-sized ($0.5 - 1.4\,R_\oplus$) planetary occurrence rate is 0.51 planets per star for orbital periods less than 50 days, significantly higher than the 0.26 planetary occurrence rate found by \citet{Petigura2013} for Earth-sized planets around solar-type stars with orbital periods between 5-100 days. A large number of studies have since explored planet occurrence as a function of orbital period, planet size and stellar spectral type using the Kepler sample \citep{Bryson2020, Kunimoto2020, Pascucci2019, Hsu2019, Zink2019, Garrett2018, Kopparapu2018, Mulders2018, Burke2015, Foreman2014, Dong2013, Youdin2011}.
A complicating factor for many of these studies is the presence of stellar companions to Kepler targets \citep{adams12,lillobox12,dressing14,law14,baranec16,furlan17,ziegler18}, which can have significant effects on exoplanet demographics both by biasing planet radii \citep{teske18} and through astrophysical effects such as the suppression of planet formation \citep{Huber2016}. In addition to exoplanet demographics, a number of studies have used asteroseismology of red giants to explore stellar populations in the Kepler field \citep{miglio13b,pinsonneault14,sharma16}.

A critical piece of information for Kepler exoplanet and stellar population studies is the process through which targets were selected. For example, most planet occurrence rates studies have so far assumed that the Kepler target selection function is unbiased with respect to stellar multiplicity. However, Kepler was forced to select targets, as only 200,000 stars could be observed over the course of the mission. Previous attempts to re-create the Kepler target selection method found that binary stars were selected at similar rates as single-star systems and that the MS dwarf population was underestimated \citep{Farmer2013}. However, \citet{Farmer2013} re-created the Kepler selection function using a synthetic population, and so many of the underlying assumptions and biases of the selection function remain unexplored.

The basis for the Kepler target selection was the Kepler Input Catalog (KIC), which contains physical properties and photometric data for sources in the Kepler field of view \citep{Brown2011}. The primary goal of the KIC was to distinguish cool dwarf stars from red giants, with an expected reliability rate of 98\% \citep{Brown2011}. The KIC used broadband photometry to infer stellar parameters for all of their stars, however the $\log(g)$ values were imprecise as they were only constrained by photometry using the D51 filter. Kepler selected the optimal targets for observation using the KIC, with a goal of selecting solar type stars that could host terrestrial-sized planets \citep{Brown2010}. The highest priority targets were solar-type stars where it would be possible to detect an Earth-sized planet in the habitable zone (HZ). The next criterion of the selection process was to include stars that were brighter than 14th magnitude in the Kepler passband ($Kp$). The following highest priority targets were the brightest stars where terrestrial-sized planets would be detectable in the HZ, even if they were as faint as 15 or 16 magnitudes. Finally, the criterion for detectable planets in the HZ was relaxed, allowing stars that would benefit from additional transit data to be observed. This created a list of 261,363 stars brighter than 16th magnitude in the Kepler passband, which was reduced to less than 200,000 stars due to mission constraints \citep{Brown2010}.

The recent Gaia second data release (DR2) now provides a unique opportunity to look back at the Kepler target sample and better understand its selection function \citep{Gaia2018}. In particular, Gaia DR2 has provided high-precision parallaxes for a total of 1,692,919,135 sources \citep{Bombrun2018} and includes nearly all the stars in the Kepler field of view, including those not observed by Kepler. The Gaia DR2 parallaxes can be used to vastly improve the properties of stars in the Kepler field \citep{Berger2018}. 

With Gaia DR2 we can now conduct a detailed investigation of the Kepler target selection function. In particular, we aim to (1) determine the degree to which Kepler's target selection matches the mission's priorities and (2) whether the selection of targets was biased with respect to stellar multiplicity. Understanding any potential biases in the selection function has important implications for exoplanet science and any future determinations of planetary occurrence rates using the Kepler target sample.

\section{Methodology}
\label{s.Method}
\subsection{Catalog Cross-matching}

We started with a subset of 2.4 million targets within the KIC that are located in the Kepler field of view, which we downloaded from the Mikulski Archive for Space Telescopes (MAST) \footnote{https://archive.stsci.edu/missions/kepler/catalogs/}. We focus on Kepler stars with $Kp < 16$ mag, as nearly all Kepler targets were below this threshold \citep{Brown2010}. 

As a first step, we cross-matched the KIC with Gaia DR2 to obtain Gaia information for each star in the KIC. To do this, we used the Centre de Données astronomiques de Strasbourg (CDS) cross-match \footnote{http://cdsxmatch.u-strasbg.fr}. This service is provided by the Université de Strasbourg and joins any VizieR data, in this case Gaia DR2, with a private data table based on the right ascension (RA) and declination (Dec) of the stars.

We conducted a positional match with a matching radius of 5 arcseconds. Frequently multiple Gaia stars were matched to a single Kepler ID, as the stars were located at similar RA and Dec. We removed duplicates by selecting only the Kepler and Gaia ID's associated with the most similar magnitudes in the Kepler passband $Kp$ and Gaia passband ($G$). 

We then extracted Gaia Re-normalized unit weight error (RUWE) values for all sources. The unit weight error (UWE) values are a representation of the normalized chi-squared values resulting from the fitting of Gaia DR2 sources to single-star point spread functions (PSFs). The RUWE value corresponds to a PSF fitting corrected for color-dependent biases. RUWE values center around 1.0, but can be large if the fit is not good or there is more noise than expected. A large RUWE value, such as 1.2 or higher, indicates a multi-stellar system where the presence of stellar companions increases the noise. RUWE values above 1.2 have been shown to be indicators of binaries that are closer than the typical $\sim$1" resolution limits of Gaia \citep{Evans2018}.

\subsection{Downselection of targets on the Kepler CCDs}

\begin{figure}
    \centering
    \includegraphics[width=0.4\textwidth]{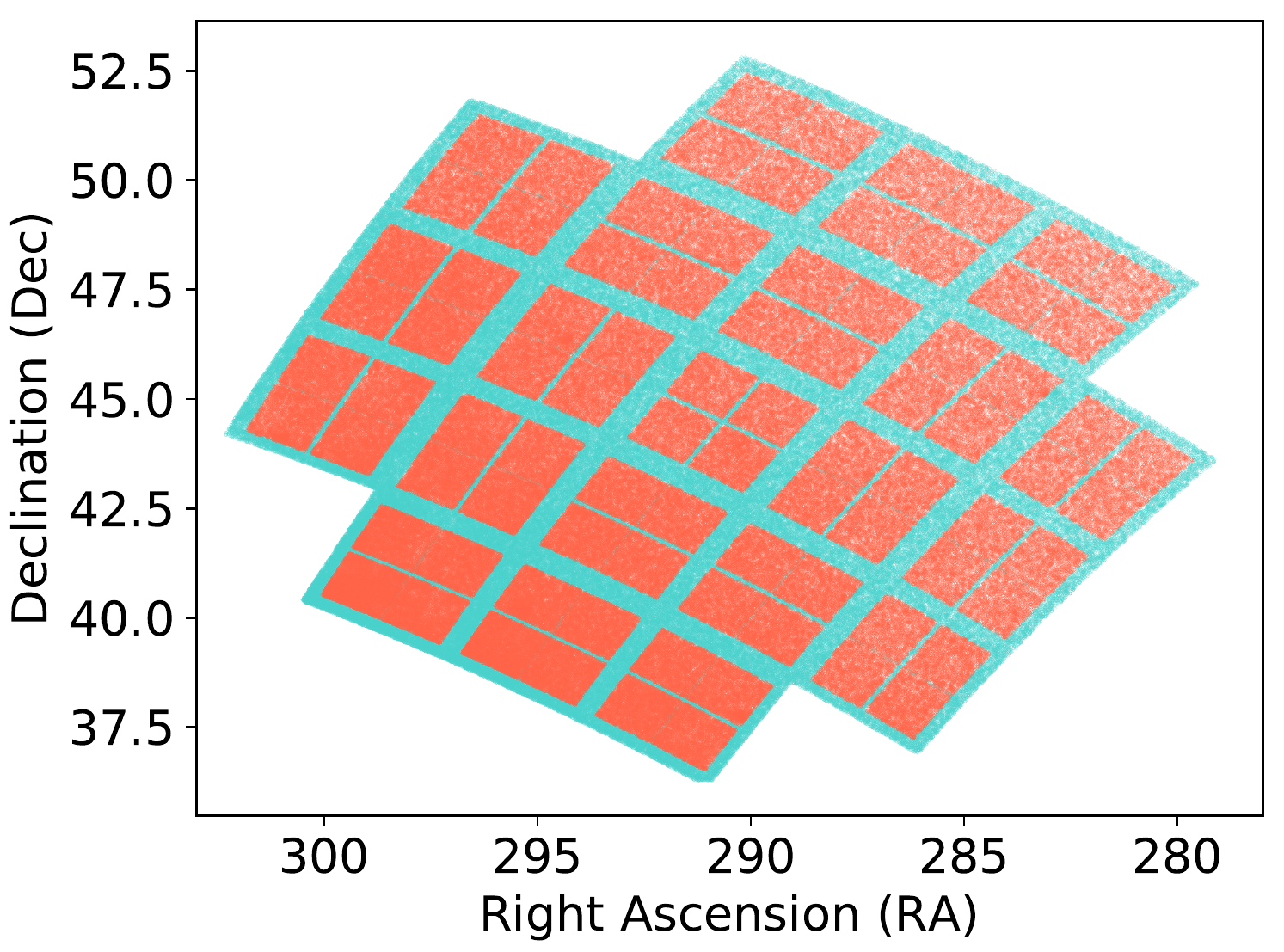}
    \caption{Spatial distribution of stars in the Kepler field. Turquoise points are all stars in the KIC near the Kepler CCDs, while red points are stars whose light did fall on the CCDs for all four seasons and are 8 arcseconds away from the CCD edges.}
    \label{fig.ra&dec}
\end{figure}

The Kepler telescope required a 90 degree roll about its optical axis every three months -- the length of a Kepler quarter -- to keep its solar array pointed at the Sun. Because these rolls produced pointing discrepancies, we define observable stars as those located more than eight arcseconds from the edges of the CCDs (two Kepler pixels) for all four Kepler quarters \citep{Borucki2010}.
Figure \ref{fig.ra&dec} shows the spatial extent of the Kepler field for stars brighter than $Kp = 16$ mag, consisting of 580,000 stars. Of the 580,000 stars in the Kepler field of view, only 379,000 stars actually fell upon the Kepler charge-coupled devices (red points). The remaining 214,000 stars' fell in the cracks of the CCDs and were physically unobservable (blue points). 
Finally, we cross-matched the dataset of observable stars with the Kepler target list \citep{Brown2010}, a table possessing the Kepler ID's of the 208,712 stars observed by Kepler. We flagged all stars without matches as non-observed stars.

\begin{figure}[hbt!]
    \centering
    \includegraphics[width=0.4\textwidth]{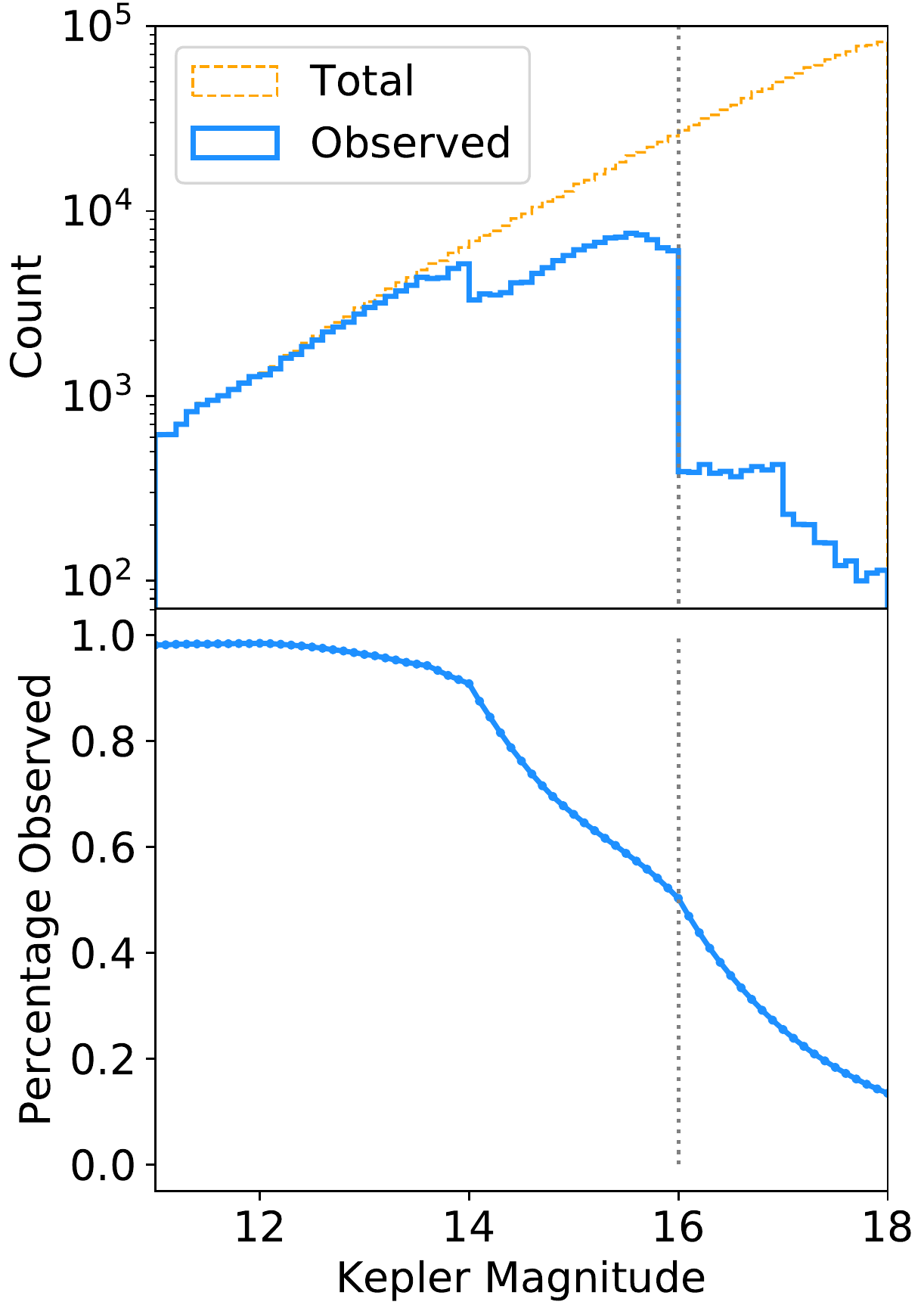}
    \caption{(a) Kepler magnitude for observed and non-observed samples. Kepler observed 194,859 stars out of the 1,559,884 stars brighter than $Kp = 18$ mag on the CCDs. (b) Percentage observed versus Kepler magnitude. The dotted grey line marks $Kp = 16$ mag.}
    \label{fig.kepmag_hist}
\end{figure}

Figure \ref{fig.kepmag_hist} shows the distributions of stars observed as a function of $Kp$ mag. The features in the histogram reflect the Kepler target selection function \citep{Brown2010}: at $Kp = 14$ mag the observed count drops, and likewise beyond $Kp = 16$ mag very few stars are observed. Stars fainter than $Kp = 16$ were selected based on more complex criteria and are not indicative of the general selection function and its biases. Therefore, we focus predominantly on stars brighter than $Kp < 16$ mag in Sections \ref{s.fullSample} and \ref{s.hostMatch}, and those with $Kp > 16$ mag in Section \ref{s.GaiaComparison}.

We calculated revised stellar parameters for all stars brighter than $Kp = 16$ mag following the method of \citet{Berger2020}. We made this decision upon discovering that over 60,000 of our 379,000 stars lacked radii and luminosity values in the Gaia archive. In addition, Gaia DR2 does not account for interstellar extinction, which can affect our data strongly as many stars in the Kepler field of view are located in the galactic plane. We used {\UrlFont{isoclassify}} \citep{Huber2017} based on the grid model from \citet{Berger2020} with Gaia parallaxes modified by the zero point offset of \citet{Lindegren2018}, Gaia G, $B_p$, and $R_p$ photometry with uncertainties to derive revised stellar parameters for all 379,000 stars. This dataset gave us access to uniform temperatures, radii, and luminosities for all stars. We subsequently reduced our dataset to 327,849 stars by removing stars with parallax uncertainties larger than 20\%. All properties are shown in Table \ref{table.values}.

\subsection{Validation of Stellar Parameters}
\label{s.comparison}

For all stars in common between our dataset and that of \cite{Berger2020}, we compared stellar effective temperatures and radii. We found no systematic offset and a $\sim$2\% scatter in our effective temperatures, and a $\sim$1\% systematic offset and a $\sim$4\% scatter in stellar radii. This scatter roughly matches the median catalog uncertainties determined in \cite{Berger2020}. As a function of both effective temperature and stellar radius, no strong trend exists in the differences between the two catalogs. Therefore, we are confident in the accuracy and precision of our derived stellar parameters.

\section{Full Kepler Sample}
\label{s.fullSample}
\subsection{HR Diagram}
\label{s.HRdiagram}

\begin{figure*}[hbt!]
    \centering
    \includegraphics[width=0.95\textwidth]{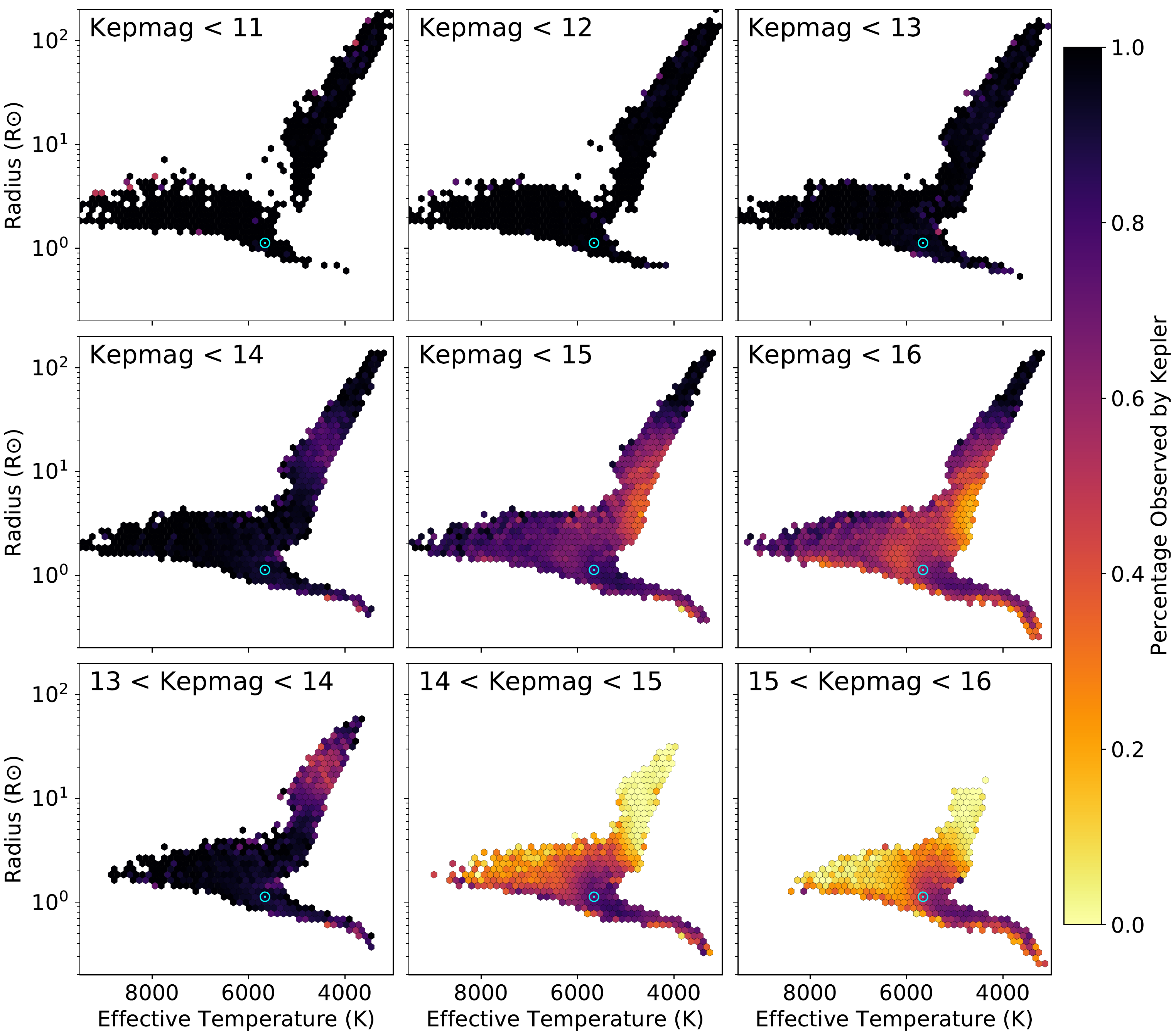}
    \caption{Stellar radius versus effective temperature for stars on the Kepler CCDs. The first 6 panels are for cumulative Kepler magnitudes and the last 3 panels are for binned Kepler magnitudes. The color of each bin corresponds to the percentage of stars observed by Kepler in that bin. The Sun is shown as the teal circled dot in each panel.}
    \label{fig.teffRad_allKmags}
\end{figure*}

 Figure \ref{fig.teffRad_allKmags} displays our derived radii and effective temperatures for subsets of the Kepler data with increasing upper limits of Kepler magnitude. The color corresponds to the percentage of stars that were observed for each effective temperature and radius bin. We observe that for Kepler magnitudes brighter than 14, nearly all stars were observed. This matches the selection function detailed in \citet{Brown2010}, as Kepler had the capacity to observe all stars brighter than $Kp < 14$ mag. 
 
 At fainter magnitudes, the HR diagrams show parameter-dependent patterns. We observe a strong selection bias against cool, low-luminosity red giants with $Kp > 14$ mag. We suspect that this is due to the fact that these stars could be more efficiently distinguished from cool dwarfs at the same temperature. Dwarfs are far more likely to host planets than red giant stars, and so the giants in this region were dropped from the target list. The observed fraction on the red giant branch is highest for the most luminous giants. This is most likely because these large giants have long pulsation periods that required the full 4 years of Kepler data to resolve \citep{Kiss2013, Yu2020, Stello2014} and because a significant number of cool giants were misclassified as giants \citep{Mann2012}.

The main sequence is well observed for bright Kepler magnitudes, but decreases substantially at $Kp = 16$ mag. We see little difference in the observed fraction between the solar-type stars and their subgiant neighbors.  This is likely because the KIC's broadband photometry was insensitive to the slight difference in $\log(g)$ values between subgiant and solar-type main-sequence stars \citep{Gaidos2013, Everett2013, Verner2011}. As a consequence, many subgiant stars were observed because they were thought to be solar-type main sequence stars, and many solar-type stars were not observed because their evolutionary state was unknown. The three bottom panels of Figure \ref{fig.teffRad_allKmags} support this claim, as we can see Kepler's broad selection of all the stars of a given temperature, whether they be subgiant or solar-type stars.

\subsection{Evolutionary States}
\label{s.evolState}

\begin{figure*}[hbt!]
\centering
\includegraphics[width=0.75\textwidth]{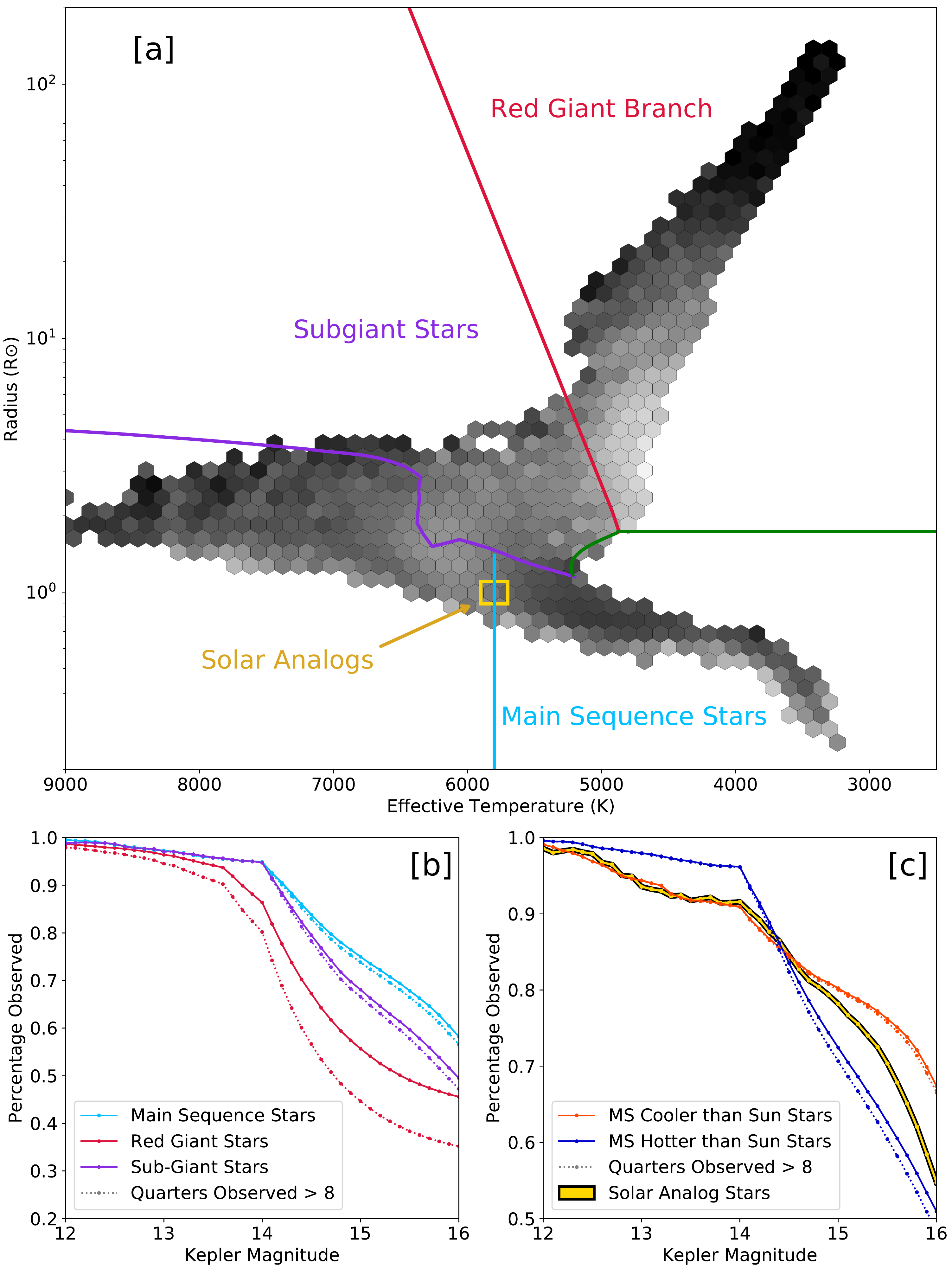} 
\caption{(a): Radius versus effective temperature for stars with $Kp < 16$ mag. The shading of each bin corresponds to the percentage of stars observed by Kepler in that bin. The lines separate and define evolutionary state/stellar property ranges. Red separates the giant and subgiants, magenta the subgiant and main sequence (MS) branch, blue the MS stars cooler and hotter than the Sun, and yellow the solar analogs ($5700$ K $< T_{eff} < 5900$ K, $0.9 < R_\odot < 1.1$). (b) and (c): Percentage observed at each $Kp$ magnitude for different evolutionary states, as defined in (a). Dotted lines represent stars observed for more than 8 quarters of the Kepler mission.}\label{fig.percentageObs}
\end{figure*}

To quantify the percentage of observed stars as a function of evolutionary state we use {\texttt{evolstate}}\footnote{http://https://github.com/danxhuber/evolstate}. These classifications assume  solar-metallicity isochrones, which on average is adequate for the Kepler field \citep{Dong2014}. \texttt{evolstate} places each star, according to its effective temperature and radius, into one of three evolutionary states: main sequence, subgiant, and giant. We additionally define solar analogs as stars with $T_{eff} = 5700-5900$\,K and $R_\odot = 0.9-1.1$. Figure \ref{fig.percentageObs}a shows an H-R diagram of our sample with the delineation of these evolutionary states marked with solid lines.

The bottom panels of Figure \ref{fig.percentageObs} show the percentage observed as a function of Kepler magnitude, colored by evolutionary state. The dotted lines of these panels are stars observed for more than 8 quarters, or one half, of the Kepler mission. Similarly, the solid lines are stars observed at any point in the mission. Figure \ref{fig.percentageObs}b confirms the conclusion of Figure \ref{fig.teffRad_allKmags} that Kepler observed nearly all stars brighter than $Kp = 14$ mag. The main sequence is the most observed evolutionary state of Figure \ref{fig.percentageObs}b, dropping to 60\% observed at $Kp = 16$ mag. The subgiant stars closely resemble the main sequence stars, although the cumulative observed percentage drops to a lower 50\% at $Kp = 16$ mag. This is most likely due to the Kepler selection function's inability to distinguish subgiant stars from solar-type main sequence stars, and it has been shown that the Kepler mission preferentially selected subgiant stars for observation \citep{Huber2014}.

From $Kp = 14-15$ mag, the fraction of observed red giants drops steeply from $\sim$80\% to $\sim$50\%, with only $\sim$ 40\% of red giants at $Kp = 15$ mag being observed for more than 8 quarters. The large separation of the red dotted line from the red solid line in Figure \ref{fig.percentageObs}b is most likely because the goal of the mission was to observe solar-type stars, and as a result many red giants were dropped from the target list after being observed for one quarter.

Figure \ref{fig.percentageObs}c breaks the main sequence into three subsections: solar analogs, MS stars cooler than the Sun, and MS stars hotter than the Sun. At $Kp = 16$ mag, the fraction of observed stars cooler than the Sun is $\sim$65\%, and the fraction of observed solar analog stars drops steeply to $\sim$55\%. We suspect this $\sim$10\% difference in observation percentages is because the small, cool dwarf stars could more easily be distinguished from giants. In addition, \citet{Dressing2013} showed that M dwarfs host a lot of small planets, which in turn led to many M dwarfs being added to the target list during later stages of the Kepler mission. In summary, our analysis shows that Kepler successfully targeted 90\% of all solar analogs brighter than $Kp < 14$ mag, decreasing to $\sim$80\% at $Kp < 15$ mag, and $\sim$55\% at $Kp < 16$ mag.

\subsection{Effective Temperatures}

\begin{figure*}[hbt!]
    \centering
    \includegraphics[width=0.95\textwidth]{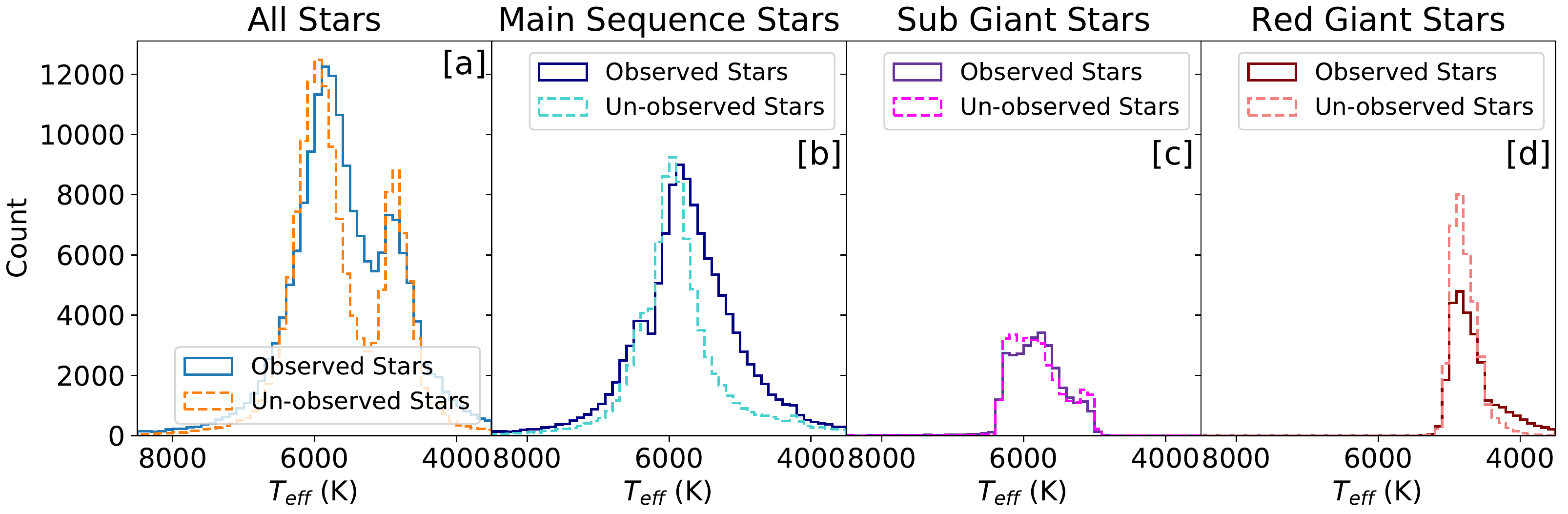}
    \caption{Histogram of the effective temperatures of Kepler stars for different evolutionary states. Solid lines represent the observed sample, and dashed lines represent the non-observed sample.}
    \label{fig.teffHist}
\end{figure*}

Figure \ref{fig.teffHist} shows histograms of the observed and non-observed samples separated by evolutionary state. As anticipated, in Figure \ref{fig.teffHist}a the observed sample peaks around solar temperature, confirming that Kepler prioritized the observation of solar-type stars. The second peak corresponds to red giants, which were mainly observed to perform asteroseismology. 

In Figures \ref{fig.teffHist}b and c, we see that for both the main sequence and subgiants the curves peak around 5800 K. This is because, as discussed previously, it was hard for Kepler to distinguish between main sequence and subgiant stars around the solar temperature. As such, all the stars with temperatures close to $T_\odot$ were observed without knowledge of their evolutionary states. For main sequence stars with temperatures below 5800 K, the percentage of stars observed is larger than it is around the solar temperature. For subgiant stars the curves are similar for all $T_{eff}$. 

Finally, in Figure \ref{fig.teffHist}d the red giant non-observed curve is significantly larger than the observed curve. This suggests that the red giant stars were not as well observed as stars of other evolutionary states, a conclusion supported by Figure \ref{fig.teffRad_allKmags} and the drop of the red line in Figure \ref{fig.percentageObs}b.

\subsection{Kinematics}
\label{s.kinematics}

\citet{McTier2019} investigated galactocentric velocities of Kepler host stars and found the host stars to be moving significantly slower than the rest of the Kepler targets. Further analysis showed that this difference was due to a selection bias in Kepler host sample, leading to the conclusion that Kepler planet occurrence is independent of galactocentric velocity. Here, we investigate whether this conclusion also holds for Kepler targets with respect to the background population. 

To do this we used our derived distances to convert Gaia DR2 proper motions into RA and Dec space motions in units of $\mathrm{kms}^{-1}$, and then added these space motions in quadrature to determine stellar sky-plane velocities. Radial velocities were only available for a small subset of stars and so we do not include them in our stellar sky-plane velocities. We reduced our investigation of kinematics to Kepler magnitudes between 14 and 16, as this is where the Kepler team was forced to make selection decisions.

\begin{figure*}[hbt!]
    \centering
    \includegraphics[width=0.95\textwidth]{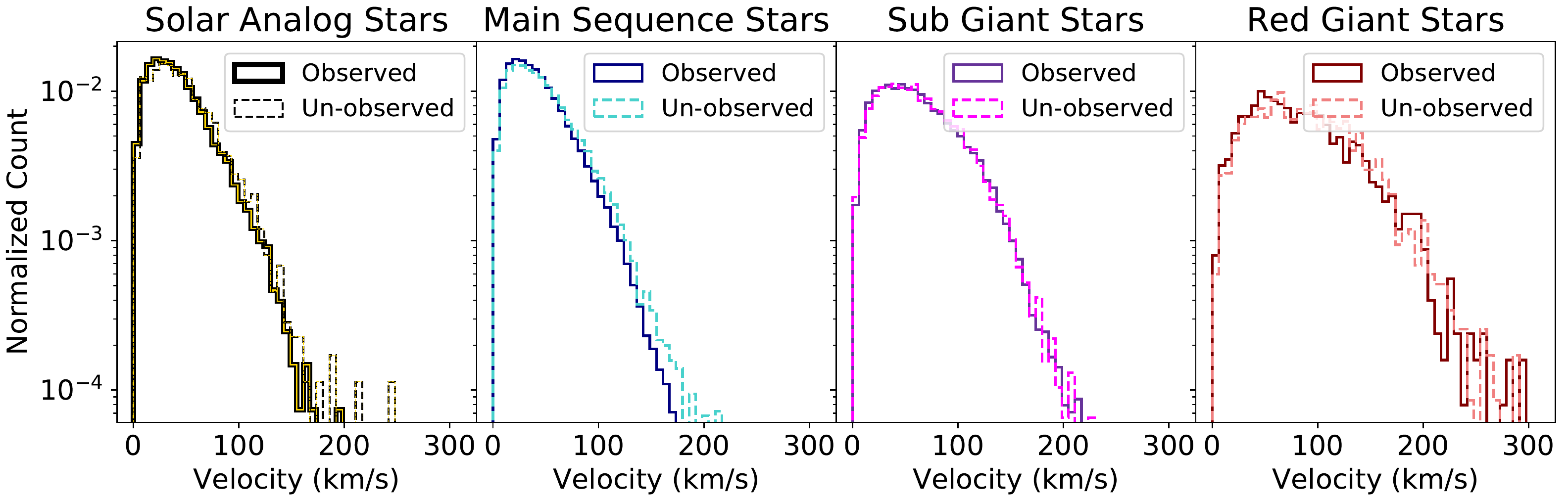}
    \caption{Histogram of the quadrature sum of the proper motions in RA and Dec ($\mathrm{kms}^{-1}$)  of Kepler stars for different evolutionary states. Solid lines represent the observed sample, and dashed lines represent the non-observed sample.}
    \label{fig.velocityHist}
\end{figure*}

The different distance distributions of the observed and non-observed samples have a strong effect on the proper motions of the samples, and therefore their sky-plane velocities. To account for this we match the non-observed sample distance distribution to the observed sample individually for each evolutionary state to the best of our ability. The distributions of sky-plane velocity for each evolutionary state is shown in Figure \ref{fig.velocityHist}.

The solar analog and main sequence stars are the only panels of Figure \ref{fig.velocityHist} showing noticeable differences between the observed and non-observed samples. The non-observed samples possess more stars with large sky-plane velocities. We attribute these differences to imperfect matches of the distance distributions of both samples, with the non-observed stars systematically extending to larger distances and thus larger velocities. 

The subgiant and red-giant distributions extend to higher velocities, as expected for stars at larger distances and similar proper motions, and from differences in kinematics for stars in different galactic populations such as the thick disc \citep{Fuhrmann1998}. In summary, we conclude that the Kepler target selection function appears to unbiased with respect to kinematics, supporting the conclusions by \citet{McTier2019} that Kepler planet occurrence is unbiased with respect to galactocentric velocities.

\subsection{Stellar Multiplicity}
\label{s.stellarMultiplicity}

Understanding biases in the target selection is important for studies investigating the effects of stellar multiplicity on planet formation using Kepler. For example, \citet{Huber2016} used AO imaging of the Kepler host star sample to conclude that the planet occurrence rate in close binary systems ($\lesssim$0.1'', $\lesssim$50\,AU) is $\sim$70\% lower than that of wider ($\sim\,0.1-1$'', $\sim$\,50-500\,AU) binaries, and thus a fifth of all solar-type stars in the Milky Way are disallowed from hosting planetary systems due to the influence of a binary companion. While the differential suppression factor derived by \citet{Huber2016} is robust against target selection bias since both close and wide binaries (as defined above) are unresolved in the KIC, the absolute scale of planet formation among binaries would be affected if there are target selection effects with respect to stellar multiplicity.

To investigate potential biases in the Kepler target selection with stellar multiplicity, we used Gaia RUWE values of the stars in the Kepler field of view for both the observed and non-observed samples. The RUWE value, as discussed in Section \ref{s.Method}, is the Gaia re-normalized unit weight error. A RUWE value above 1.2 suggests that the star in question has a binary companion. Thus, if the Kepler observed and non-observed samples have different distributions for large RUWE we can conclude that the selection function was biased in some way with respect to stellar multiplicity.

Similarly to the analyses of Section \ref{s.kinematics} we first reduced our sample to the stars with $14 < Kp < 16$ mag, both because the selection function was most active in this magnitude range and because the unit-weight error (UWE) has been shown to be inaccurate for Gaia G magnitudes brighter than 13th mag \citep{Lindegren2018}. As a next step we matched the non-observed sample distance distribution to the observed sample distribution individually for each evolutionary state. This is because RUWE values are dependent on the apparent angular separation of binary companions, and the intrinsic physical separations of binary companions will produce different RUWE distributions depending on how far away that system is from Earth. The RUWE distributions are shown below in Figure \ref{fig.ruweHist}.

\begin{figure*}[hbt!]
    \centering
    \includegraphics[width=0.95\textwidth]{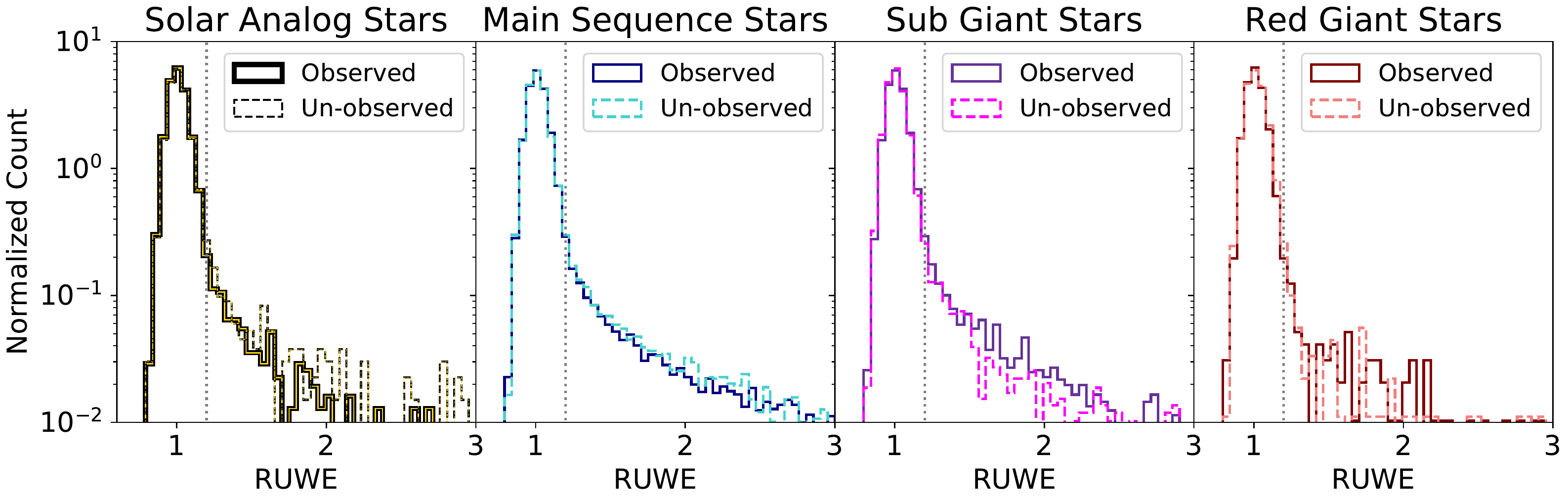}
    \caption{Histogram of RUWE values of Kepler stars for different evolutionary states. Solid lines represent the observed samples, and dashed lines represent the non-observed samples. The dotted grey line marks the RUWE value beyond which stars are likely to possess companions.}
    \label{fig.ruweHist}
\end{figure*}

RUWE values greater than 1.2 (shown by the dotted grey line in Figure \ref{fig.ruweHist}) correlate with multiplicity, and hence we focus we focus our investigation on the stars with large RUWE values. To quantify the fraction of stars with stellar companions, we divided the number of stars with RUWE $> 1.2$ by the total number of stars. We repeated this process for both the observed and un-observed samples for various evolutionary states. 

The red giant observed and non-observed samples have moderately significant differences and match within $3 \sigma$. In contrast, the main sequence, solar analog, and subgiant observed and non-observed samples differ by $\gtrsim 4 \sigma$. While this difference is significant it is important to note that stellar multiplicity, and thus the RUWE values, can also induce significant biases in the evolutionary state classifications themselves by affecting the derived luminosity values. Specifically, the inverse relationship between the samples is likely due to Malmquist bias, which causes main sequence stars in binary systems to appear as subgiants due to their inflated luminosity values. This would cause the main sequence observed sample to have a lower fraction of binary stars than the non-observed sample, because the observed binaries in the main sequence have been misidentified as subgiant stars. Similarly, it would cause the subgiant observed sample to have a greater fraction of binary stars than the non-observed sample since the observed sample include main sequence binary stars as well as subgiant binary stars. We therefore attribute the differences in the RUWE distributions in Figure \ref{fig.ruweHist} to effects of binaries on our derived stellar parameters, highlighting the importance of calibrating Malmquist bias when using Gaia to assess the impact of binaries on planet occurrence (Kraus et al., in prep).

\section{Host Star Sample}
\label{s.hostMatch}

\begin{figure*}[hbt!]
    \centering
    \includegraphics[width=0.8\textwidth]{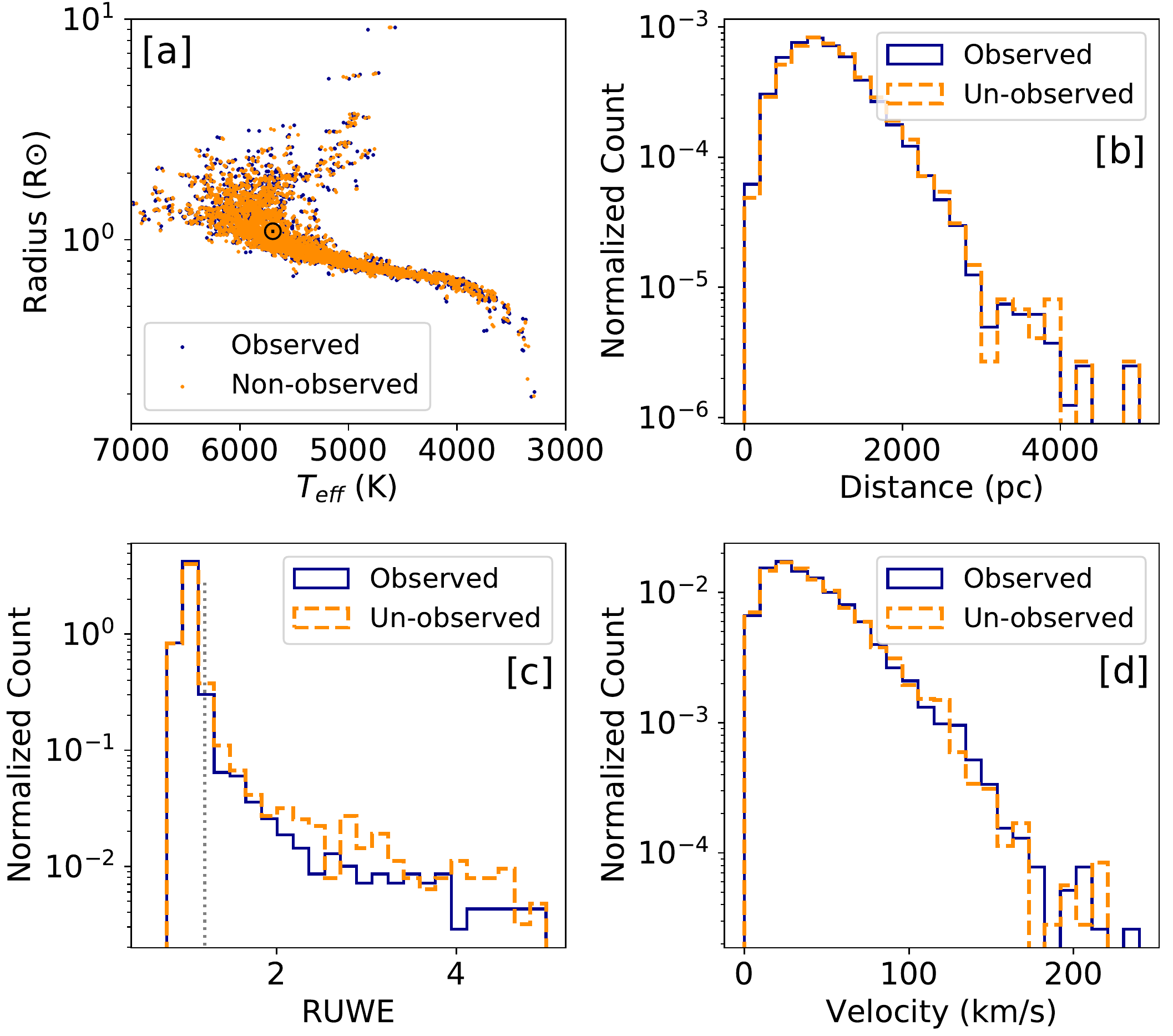}
    \caption{(a) HR diagram of the observed (blue) and non-observed (orange) samples. (b) Histogram of the distance distributions for the observed and the non-observed samples. (c) Histogram of the RUWE values for the observed and the non-observed samples. (d) Histogram of the sky-plane velocity distributions for the observed and the non-observed samples.}
    \label{fig.hostMatched}
\end{figure*}

To better control for the biases entering the kinematic and multiplicity analysis for the full sample, we performed a separate analysis focusing only on the sample of stars with transiting planets. We used the Kepler host sample as a basis for our investigations into the differences between the observed and non-observed samples RUWE and sky-plane velocity distributions. Our host sample consists of 2,066 stars from the KOI table of the NASA Exoplanet Archive \citep{akeson13} with either confirmed or candidate planets and Kepler magnitudes fainter than 14 and brighter than 16 mag \citep{Thompson2018}. We randomly selected stars from our observed and non-observed samples that match the host sample distributions of effective temperature, radius, and distance. By matching our observed and non-observed samples to the host sample we create two samples that are similar in both evolutionary state and distance. As such, if any differences arise in our comparisons of RUWE and sky-plane velocity values between the matched samples they are due to the Kepler selection function. The comparison between the host-matched observed and non-observed samples is shown in Figure \ref{fig.hostMatched}.

Figure \ref{fig.hostMatched} (a) and (b) confirm that the distributions of effective temperature, radius, and distance are matched between the observed and non-observed samples. Figure \ref{fig.hostMatched} (c) shows the RUWE values of the observed and non-observed samples, and it can be seen that the non-observed sample is consistently above the observed sample for high RUWE values. $8.2 \pm 0.5$\% of the observed sample have RUWE values above 1.2 and $11.8 \pm 0.6$\% of the non-observed sample have RUWE values above 1.2. This significant $4.8 \sigma$ difference suggests that Kepler preferentially selected non-binary stars for observation. This result persists when taking into account Gaia DR3 values.

The difference in elevated RUWE for observed and non-observed Kepler targets may have implications on studies of planet occurrence as a function of multiplicity. To investigate whether this difference probes close or wide binaries we compared the KIC contamination numbers for both samples, which  trace effects of nearby stars that are resolved in the KIC. We found that both samples have KIC contamination numbers that match within 0.5\,$\sigma$, implying that the RUWE difference in Figure \ref{fig.hostMatched}c may be mostly driven by binaries that are unresolved in the KIC. Based on this we speculate that the target selection bias may have been produced by unresolved companions causing broadband colors which deviate from predictions from single star model atmospheres, which were used to perform the stellar classification in the KIC \citep{Brown2011}. This is further supported by the fact that the difference between the samples is largest in the moderate RUWE regime ($\sim\,2-5$), which probe higher contrast companions that would remain undetected in the seeing-limited imaging used to construct the KIC. However, we cannot rule out that some fraction of the difference in RUWE values between observed and unobserved sample can be attributed to wide companions that were intentionally removed.

In contrast to RUWE, the velocity distributions of Figure \ref{fig.hostMatched}d appear very similar, and a Kolmogorv-Smirnov test confirms this conclusion with a p-value of 0.97. This p-value, as well as the similarities in the sky-plane velocity distributions of Figure \ref{fig.hostMatched}d, allow us to conclude that Kepler was unbiased with respect to proper motions, confirming the results by \citet{McTier2019}.

\section{Comparison with Gaia DR2 Stellar Parameters}
\label{s.GaiaComparison}

\subsection{Revised Radius Comparison}
\begin{figure*}
    \centering
    \includegraphics[width=0.8\textwidth]{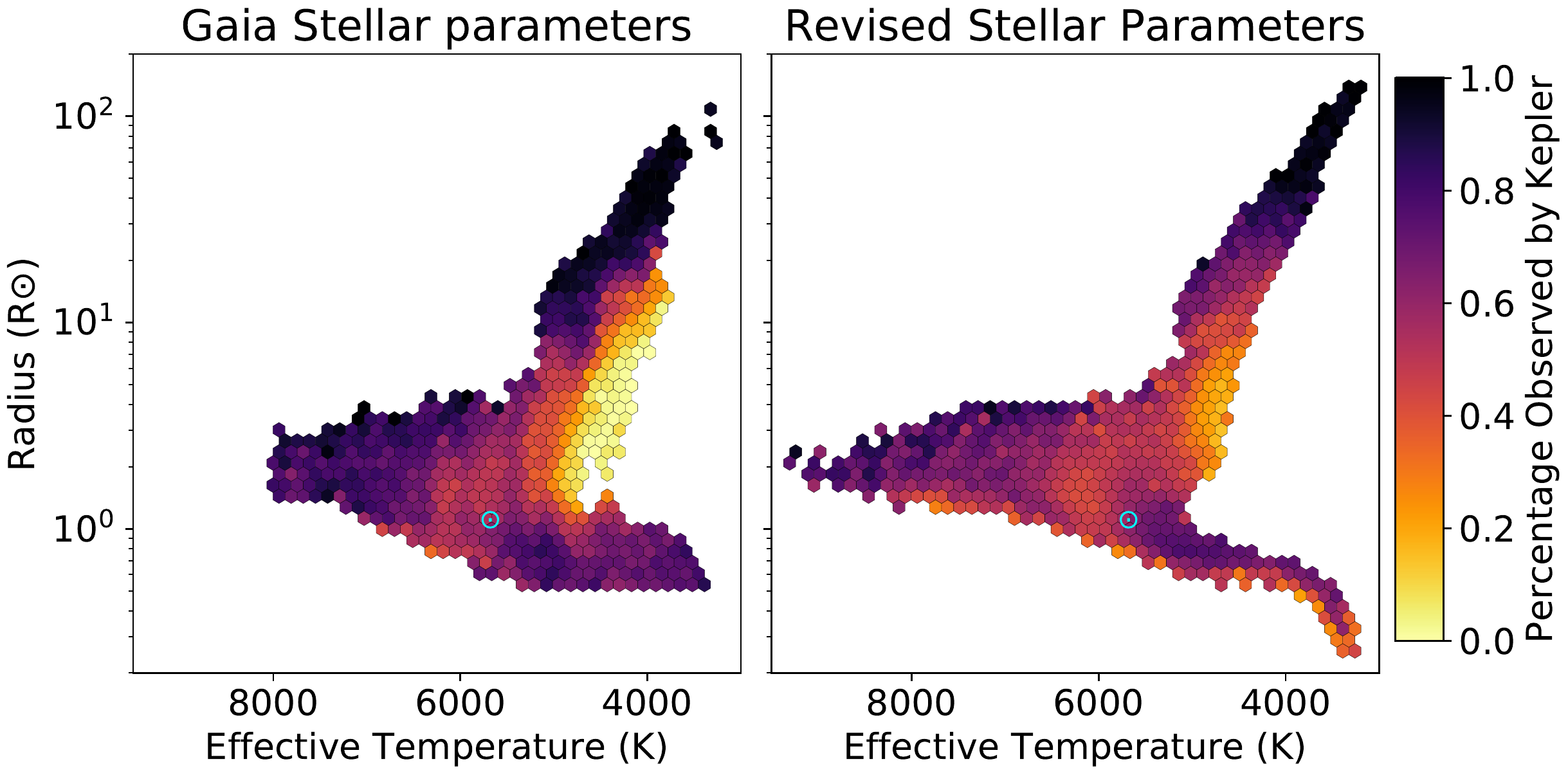}
    \caption{HR diagrams for stars brighter than $Kp = 16$ mag. Figure (a) shows the Gaia radii and effective temperatures. Figure (b) shows our revised radii and effective temperatures. The Sun is shown as the circled dot in each panel.}
    \label{fig.gaiaComparison}
\end{figure*}

In previous sections of this paper we use the revised stellar parameters calculated with the techniques of \citet{Berger2020} for our analysis. However, since these properties depend on evolutionary models, we also investigated the difference of these parameters to those provided in the Gaia DR2 archive \citep{andrae18}. The latter also allow us to investigate the properties of the faintest selected Kepler targets ($Kp > 16$ mag), which were excluded from our classifications.

Figure \ref{fig.gaiaComparison} shows a comparison between the HR diagrams of the Gaia stellar parameters and our revised stellar parameters. The diagrams of Figure \ref{fig.gaiaComparison} look qualitatively similar, confirming our main conclusions of the previous sections. The red giant branch stars show similar color variations along the radius axis, and the subgiant stars are observed at relatively the same rate as solar-type stars in both diagrams. However, Gaia DR2 is missing both the hottest and coolest main sequence stars, due to their cuts on $T_{eff}$ and luminosity (only stars with $3300 < T_{eff} < 8000$ K and $\sigma(L)/L > 0.3$ were given luminosity and $T_{eff}$ values) \citep{Andrae2018}.

\subsection{Distribution of Faint ($Kp > 16$ mag) Kepler Stars}
\begin{figure*}
    \centering
    \includegraphics[width=0.95\textwidth]{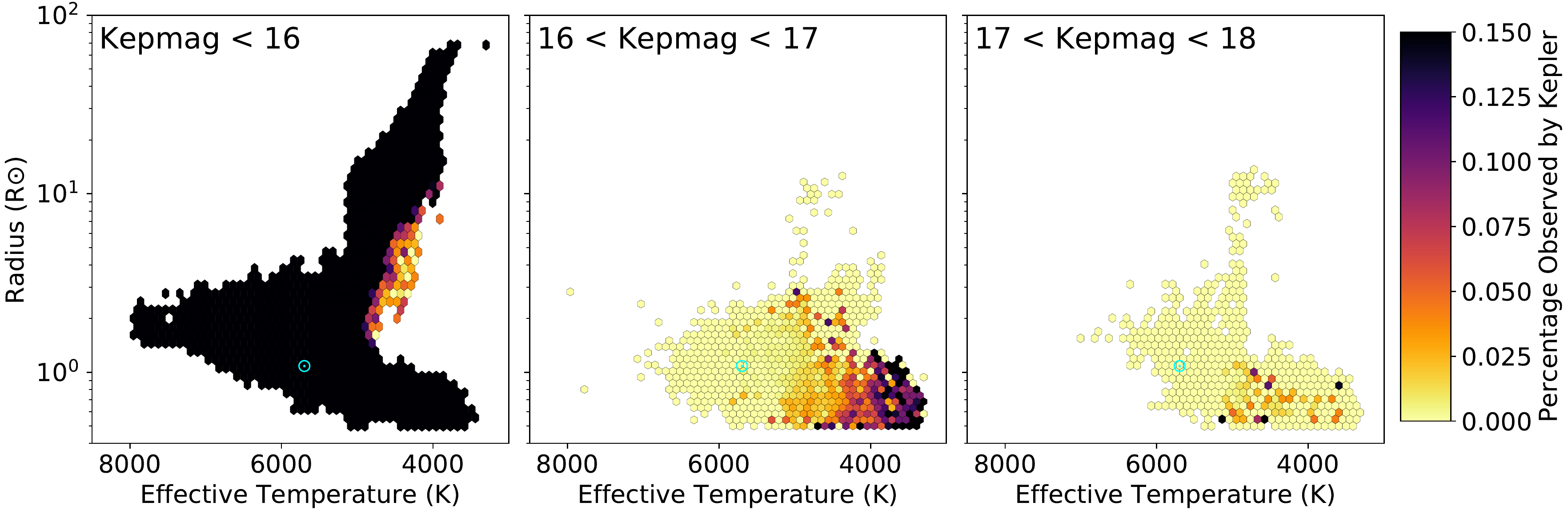}
    \caption{Gaia DR2 radii plotted against $T_{eff}$ for increasing Kepler magnitudes beyond 18th magnitude. The color of each bin corresponds to the percentage of points observed by Kepler in that bin. The Sun is shown as the circled dot in each panel. 302,008 stars are plotted in (a), 163,572 stars in (b), and 15,066 stars are plotted in (c).}
    \label{fig.teffRadius_kp18}
\end{figure*}

One advantage of using the Gaia DR2 stellar parameters is that they provide parameters for stars fainter than $Kp = 16$ mag. These faint stars are displayed in Figure \ref{fig.teffRadius_kp18}, with color-coding showing the percentage of stars observed by Kepler on a scale from 0$\%$ to 15$\%$. We observe the that vast majority of targeted stars fainter than $Kp > 16$ mag are cool dwarfs. We suspect that this is due to the fact that towards the end of the Kepler mission it was discovered that M dwarf stars host many planets \citep{Dressing2013} and were added to the Kepler target list. We see few stars with $Kp > 17$ mag because of Gaia's inability to derive physical parameters for faint stars \citep{Gaia2018}.

\subsection{Binary Fraction}

Gaia DR2 derived their radii from the Stefan-Boltzmann law. Radii derived this way (rather than using isochrones) for cool dwarfs will form a second main-sequence, which is made up of cool dwarf stars that appear more luminous on an H-R diagram than they actually are due to the presence of stellar companions. This "binary main sequence" provides us with another metric to analyze the Kepler selection function's bias with respect to stellar multiplicity.

\begin{figure*}
    \centering
    \includegraphics[width=0.75\textwidth]{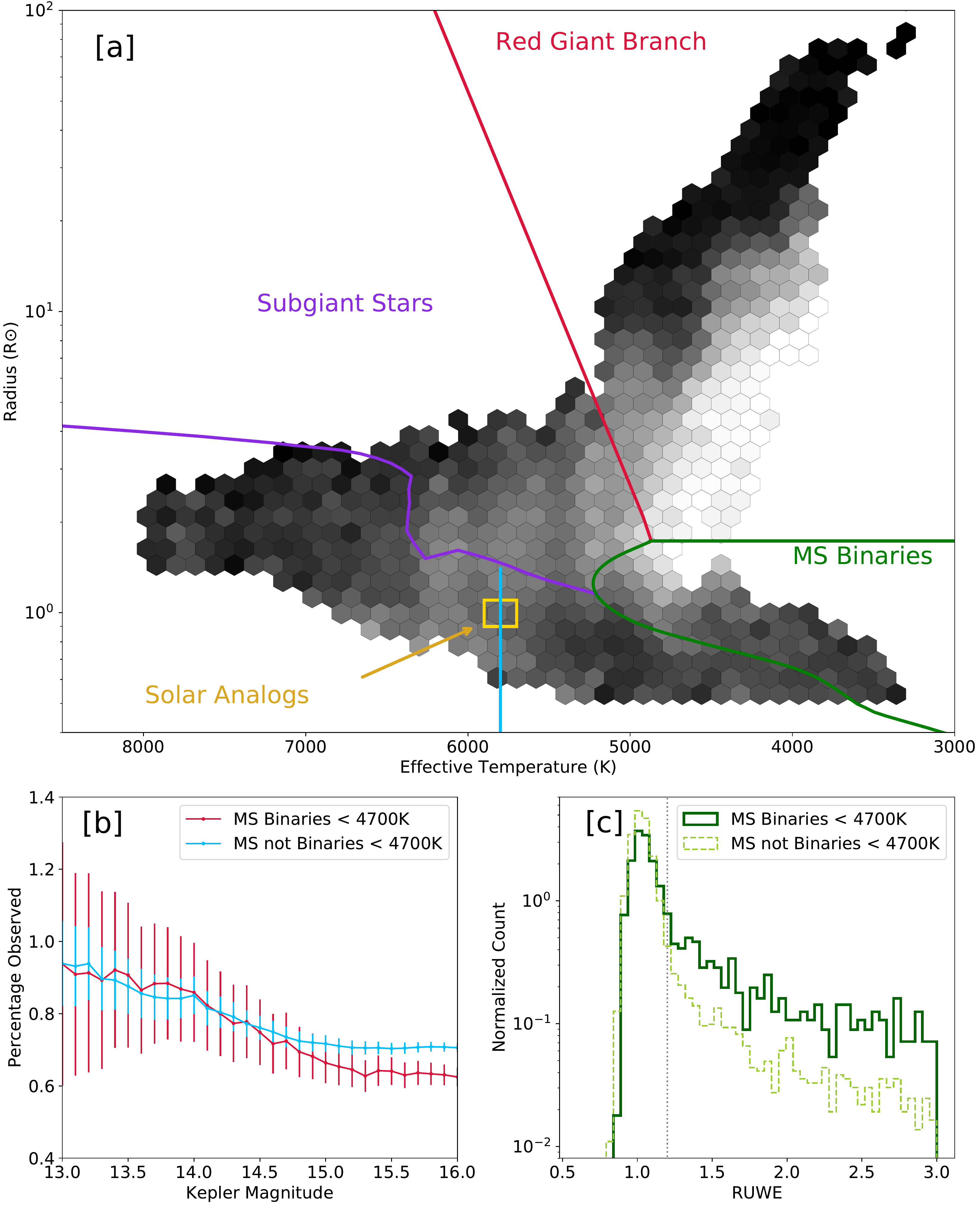}
    \caption{(a): Effective temperature and radius plot of $Kp < 16$ mag with lines defining different evolutionary states. Red separates the giant and subgiants, magenta the subgiant and main sequence stars, green the suspected main sequence binaries, blue the stars cooler and hotter than the Sun, and yellow the solar analogs. (b): Percentage observed at each $Kp$ magnitude for main sequence and main sequence binary stars with $T_{eff} < 4700$ K. (c): Histogram of the RUWE values for main sequence and main sequence binary stars with $T_{eff} < 4700$ K. The dotted grey line marks the RUWE value beyond which stars are likely to possess companions.}
    \label{fig.percentObs_binaries}
\end{figure*}

Figure \ref{fig.percentObs_binaries}a displays the same information as Figure \ref{fig.percentageObs}a, except we now plot Gaia DR2 $T_{eff}$ and radii and use the definitions of \citet{Berger2018} to identify cool main sequence binaries (green line). Figure \ref{fig.percentObs_binaries}c demonstrates that the green line in Figure \ref{fig.percentObs_binaries}a indeed efficiently identifies binaries by comparing the RUWE values of the two cool main sequence star samples. $39 \pm 2\%$ of the cool main sequence binary stars have RUWE values greater than 1.2 and $15.2 \pm 0.5\%$ of the cool main sequence stars have RUWE values greater than 1.2, leading to a difference of $12.1 \sigma$. 

Figure \ref{fig.percentObs_binaries}b displays the percentage observed of cool main sequence stars and their neighboring binaries as a function of $Kp$ magnitude. The percentage observed of cool main sequence stars increases significantly for $Kp > 14$ mag, with $\sim$8\% fewer cool main sequence binaries observed. This qualitatively agrees with the conclusion of Section \ref{s.hostMatch} that Kepler preferentially selected non-binary stars for observation. 

\section{Conclusions}

In this paper we have analyzed the Kepler mission's target selection function by using Gaia DR2 as the ground truth to characterize the $\sim$500,000 stars that Kepler could have observed, and compared this population to the sample of $\sim$200,000 stars that were selected for observations. Our main results are as follows:

\begin{itemize}
    \item{We find that the Kepler target selection was efficient at selecting solar-type stars. Specifically, the Kepler target selection is essentially complete for $Kp < 14$ mag, with the main-sequence star selection fraction dropping from 95\% to 60\% between $14 < Kp < 16$ mag. For stars on the main sequence completeness is best for stars cooler than the Sun and worst for stars hotter than the Sun, with 55\% of all solar analogs observed for $Kp < 16$ mag. We find that the observed fraction for subgiant stars is only $\sim$10\% lower than the main sequence stars, implying that many subgiant stars selected for observation were believed to be main sequence stars.} 
    \item{We find that the observed fraction for red giant stars drops from 90\% at $Kp = 14$ mag to 45\% at $Kp = 16$ mag. Kepler's selection of red giant stars was most strongly biased against cool, low luminosity giants, with completeness dropping below 30\%. This confirms that the KIC was efficient in separating giants from dwarfs, in particular for temperatures between $4000-5000$\,K.}
    \item{We find that the distribution of elevated Gaia re-normalized unit weight error (RUWE $> 1.2$) of the observed and non-observed main-sequence stars differ at $\sim$5\,$\sigma$ significance, implying a Kepler target selection bias against binaries. This conclusion is robust when taking into account differences in the sample properties and supported when using the luminosities of cool main-sequence stars as a proxy for binarity. We find tentative evidence that the RUWE difference may be caused by close binaries that were unresolved in the KIC and speculate that biases in composite broadband colors may have led to this selection bias, but further work will be needed to confirm this conclusion. The difference in elevated RUWE does not affect the previously detected differential planet formation suppression rate for close binaries \citep{Huber2016}, but highlights the importance of taking into account selection biases for determining the absolute scale of stellar multiplicity effects on planet occurrence.}
    \item{We find that Kepler target sample is unbiased with respect to galactocentric space velocities compared to the background population of stars Kepler could have selected for observations. This confirms previous results for Kepler exoplanet host stars by \citet{McTier2019}.} 
    \item{We find that the faintest Kepler stars were exclusively selected to be cool dwarfs. The observed M dwarf fraction is $\sim$14\% for $16 < Kp < 17$ mag, and falls to $\sim$8\% for $17 < Kp < 18$ mag.}
\end{itemize}

Gaia DR2 has enabled the first comprehensive evaluation of the biases and successes of the Kepler selection function, which will be valuable for the study of exoplanet demographics and stellar populations using Kepler data. For example, the bias against stellar multiplicity identified suggests that future research may require analysis of a control sample of non-observed, non-host stars. Future studies combining Gaia RUWE with AO imaging will be required to determine whether the bias identified here is caused by wide or close binaries. Additionally, future Gaia data releases with improved resolution and source completeness will allow more detailed investigations of the selection function bias for Kepler and other space-based transit surveys. We note that Gaia EDR3 was released during the final phases of completing this paper \citep{lindegren20}. We have performed basic comparisons of parallaxes, kinematics and RUWE values for the Kepler sample and confirmed that our main conclusions remain unchanged when using results from EDR3.

\section*{Acknowledgements}
We gratefully acknowledge the Gaia and Kepler missions and the people involved for their hard work that have made this paper possible. We thank Adam Kraus for discussions on stellar multiplicity bias in the Kepler sample and helpful feedback on the paper.
We also thank Robert Jedicke, Aaron Do, Ben Shappee, Michael Bottom, Victoria Catlett, Jingwen Zhang, Casey Brinkman, Ashley Chontos, Vanshree Bhalotia, Nicholas Saunders, Larissa Nofi, Jamie Tayar, and Lauren Weiss for their helpful discussions and feedback.

L.M.W. acknowledges support from Research Experience for Undergraduate program at the Institute for Astronomy, University of Hawaii-Manoa funded through NSF grant 6104374. L.M.W. would like to thank the Institute for Astronomy for their kind hospitality during the course of this project.

T.A.B. and D.H. acknowledge support from a NASA FINESST award (80NSSC19K1424) and the National Science Foundation (AST-1717000). D.H. also acknowledges support from the Alfred P. Sloan Foundation.

This work has made use of data from the European Space Agency (ESA) mission Gaia (https://www.cosmos.esa.int/gaia), processed by the Gaia Data Processing and Analysis Consortium (DPAC; https://www.cosmos.esa.int/web/gaia/dpac/consortium). Funding for the DPAC has been provided by national institutions, in particular the institutions participating in the Gaia Multilateral Agreement. This research has made use of NASA’s Astrophysics Data System. This research has made use of the VizieR catalogue access tool, CDS, Strasbourg, France. We obtain our Kepler data from MAST, and our Gaia DR2 data from VizieR. This paper has also made use of the open-source TOPCAT application.
\software{pandas, SciPy, Matplotlib}.


\bibliography{references}
\bibliographystyle{aasjournal}


\setcounter{figure}{0} \renewcommand{\thefigure}{A.\arabic{figure}}
\setcounter{table}{0} \renewcommand{\thetable}{A.\arabic{table}}
\setcounter{section}{0} \renewcommand{\thesection}{A.}
\setcounter{subsection}{0}

\section*{Appendix}

\begin{table}[h!]
 \centering
 \caption{Stellar Properties of all stars that fall on the Kepler CCDs}
 \begin{tabular}{||c c c c c c c c c c c||}
 \hline
 KIC ID & Gaia DR2 ID & obsFlag & hostFlag & Kp & $T_{eff}$ & Radius & Distance & RUWE & Velocity & evolState \\ [0.5ex]
 \hline
  & & & & & K & $R_{\odot}$ & Pc & & $kms^{-1}$ & \\
 \hline\hline
 757076 & 2050233807328471424 & 1 & 0 & 11.7 & 5135 & 4.08 & 652 & 0.947 & 44.98 & 2 \\
 \hline
 757099 & 2050233601176543104 & 1 & 0 & 13.2 & 5448 & 0.98 & 368 & 2.173 & 6.85 & 0 \\
 \hline
 891968 & 2050246795316089088 & 0 & 0 & 14.7 & 5632 & 0.99 & 816 & 1.006 & 38.16 & 0 \\
 \hline
 892010 & 2050234975566082176 & 1 & 0 & 11.7 & 4572 & 15.15 & 1826 & 1.014 & 97.71 & 2 \\
 \hline
 892107 & 2050234696381511808 & 1 & 0 & 12.4 & 4904 & 4.52 & 937 & 0.940 & 65.46 & 2 \\
 \hline
 892119 & 2050235113005074304 & 0 & 0 & 15.2 & 4830 & 6.30 & 4555 & 1.007 & 116.34 & 2 \\
 \hline
 892195 & 2050234735047928320 & 1 & 0 & 13.8 & 5371 & 0.97 & 479 & 1.122 & 18.95 & 0 \\
 \hline
 892202 & 2050236521754351360 & 0 & 0 & 15.7 & 5997 & 1.16 & 1723 & 1.014 & 13.01 & 0 \\
 \hline
 892203 & 2050236521754360832 & 1 & 0 & 13.6 & 5690 & 1.06 & 554 & 1.245 & 13.56 & 0 \\
 \hline
 892212 & 2050233876054461056 & 0 & 0 & 14.4 & 5405 & 0.88 & 1281 &  & 45.48 & 0 \\
 \hline
 \end{tabular}
 \\[13 pt]
 The first 10 rows of the dataset used for our analysis. The full table, in machine-readable format, can be found online. obsFlag: 1 is observed, 0 is not. hostFlag: 1 is a host star, 0 is not. evolState: 0 is Main Sequence, 1 is Sub Giant, 2 is Red Giant.
 
 \label{table.values}
\end{table}

\end{document}